%% file: arxiv.tex
\documentclass{IEEEtran}  

\usepackage[utf8]{inputenc}
\usepackage[footnotes,definitionLists,hashEnumerators,smartEllipses,hybrid]{markdown}
\usepackage[capitalise,noabbrev,nameinlink]{cleveref}
\usepackage{graphicx}
\usepackage{adjustbox}
\usepackage{graphbox}
\usepackage{import}
\usepackage{epstopdf}
\usepackage[labelformat=simple]{subcaption}
\usepackage{siunitx}
\usepackage{tabularx}
\usepackage{booktabs}
\usepackage{colortbl}
\usepackage{fp}
\usepackage{tikz}
\usetikzlibrary{calc}
\DeclareCaptionFormat{overlay}{\gdef\capoverlay{#1#2#3\par}}
\DeclareCaptionStyle{overlay}{format=overlay}
\DeclareCaptionFormat{suboverlay}{\gdef\subcapoverlay{(\thesubfigure) #3\par}}
\DeclareCaptionStyle{suboverlay}{format=suboverlay}
\usepackage{tikzscale}
\usepackage{pgfplots}
\usepgfplotslibrary{groupplots}
\pgfplotsset{width=10cm,compat=1.16}
\usepackage{svg}
\usepackage{todonotes}

\usepackage[numbers]{natbib}

\definecolor{goodgreen}{HTML}{c5eecc}
\definecolor{goodred}{HTML}{ffc7ce}
\newcommand{\MinNumber}{-1}%
\newcommand{\MaxNumber}{1}%

\newcommand{\ApplyGradient}[1]{%
  \FPeval{\%}{100.0*(#1-\MinNumber)/(\MaxNumber-\MinNumber)}%
  \pgfmathsetmacro{\%Color}{\%}%
  \xdef\%Color{\%Color}%
  \cellcolor{goodgreen!\%Color!goodred}{#1}%
}

\newcolumntype{H}{>{\collectcell\ApplyGradient}X<{\endcollectcell}}

\DeclareCaptionFormat{overlay}{\gdef\capoverlay{#1#2#3\par}}
\DeclareCaptionStyle{overlay}{format=overlay}

\title{\huge
{Influence of Rotator Cuff Integrity on Loading and Kinematics\\ Before and After Reverse Shoulder Arthroplasty}
}

\author{Fabien P\'{e}an, Philippe Favre, Orcun Goksel
\thanks{F.~P\'{e}an and O.~Goksel are with the Computer-assisted Applications in Medicine (CAiM) group, ETH Zurich, Switzerland.}
\thanks{P.~Favre is with Zimmer Biomet, Winterthur, Switzerland.
}}

\begin{document}

\maketitle
\begin{abstract}
Reverse Shoulder Arthroplasty (RSA) has become a very common procedure for shoulder joint replacement, even for scenarios where an anatomical reconstruction would traditionally be used.
In this study, we investigate joint reaction forces and scapular kinematics for
rotator cuff tears of different tendons with and without a reverse prosthesis.
Available motion capture data during anterior flexion was input to a finite-element musculoskeletal shoulder model, and muscle activations were computed using inverse dynamics.
The model was validated with respect to in-vivo glenohumeral joint reaction force (JRF) measurements, and also compared to existing clinical and biomechanical data.
Simulations were carried out for the intact joint as well as for various
tendons involved in a rotator cuff tear: superior (supraspinatus), superior-anterior (supraspinatus and subscapularis), and superior-posterior (supraspinatus, infraspinatus and teres minor). 
Each rotator cuff tear condition was repeated after shifting the humerus and the glenohumeral joint center of rotation to simulate the effect of a reverse prosthesis.
Changes in compressive, shear, and total JRF were analysed, along with scapular upward rotation.
The model compared favourably to in-vivo measurements, and existing clinical and biomechanical knowledge. Simulated JRF lie in the ranges of in-vivo JRF measurements and shows a linear increase past 90\si{\degree} flexion. 
Implanting a reverse prosthesis with a functional rotator cuff or with an isolated supraspinatus tear led to
over 2 times higher compressive force component than with massive rotator cuff tears (superior-anterior or superior-posterior).
Such higher compression forces might increase the risk of wear and implant fracture.
\end{abstract}

\section{Introduction}
Shoulder arthroplasty is a common clinical procedure to treat arthropathy, osteoarthritis or rotator cuff tears~\cite{Drake2010,Ricchetti2011}.
Anatomical arthroplasty is generally indicated for patients with a functional rotator cuff, while reverse shoulder arthroplasty is primarily intended for patients with massive rotator cuff tears (MCRT)~\cite{Flatow2011,Mattei2015}.
Unlike the anatomical version, the reverse procedure involves placing a socket-shaped implant in the humerus shaft and a ball-shaped base-plate on the glenoid, which forms a stable ball-socket joint thanks to the enclosing socket.
Such reverse arthroplasty ball-socket glenohumeral joint typically has a center-of-rotation slightly offset to the corresponding physiological case~\cite{Boileau2005}.
In practice, reverse prostheses are often easier to implant, and subject to less loosening over the long term ~\cite{Casagrande2016,Gregory2017}.
Moreover, rotator cuff tears are a common complication of anatomical shoulder arthroplasty, and revision surgery generally consists of converting the anatomical into a reverse reconstruction~\cite{Walker2012,Abdel2013,Alentorn-Geli2015,Hernandez2017}.
The revision procedure is difficult, as the cemented polyethylene glenoid component needs to be removed, sacrificing glenoid bone and putting the support condition of the reverse baseplate at risk.
Consequently, an increasing number of patients with a functional rotator cuff are directly treated with a reversed shoulder arthroplasty, even when an anatomical reconstruction is indicated~\cite{Palsis2018}.

The effect of implanting a reverse prosthesis with a near-functional rotator cuff remains unclear, considering that  reverse prostheses have been originally designed to compensate for the missing rotator cuff function.
Our study aims to investigate the impact of various rotator cuff tear conditions with a reverse implant and its potential long-term implications.
Using a previously published musculoskeletal model of the shoulder, we simulated active motion and evaluated the glenohumeral joint reaction forces~(JRF) and scapular kinematics for superior, superior-anterior, and superior-posterior tears of the rotator cuff, with and without a reverse implant.

\section{Methods}
\subsection{Musculoskeletal Model of the Shoulder}
The biomechanical model used herein, developed upon open-source biomechanical framework Artisynth~\cite{Lloyd2012}, has been previously described in detail~\cite{Pean2020surface}, but the main relevant features are summarized below for sake of completeness.

Our multi-body model involves four rigid structures: thorax, clavicle, scapula, humerus.
With the thorax fixed in space, the resulting kinematics chain consists of three spherical joints: sternoclavicular between the thorax and the clavicle, acromioclavicular between the clavicle and the scapula, and glenohumeral between the scapula and the humerus.
The acromioclavicular and coracoclavicular ligaments are each represented by a spring (with a stiffness of $10^6\,$N/m) located at the volar and dorsal extremities of the acromion and the coracoid processes, respectively.
The scapulo-thoracic pseudo-joint is maintained by collision detection and contact mechanics.

We automatically generated B-spline surface meshes~\cite{Pean2020surface} for 19 muscles segments representing 13 muscles, which were modelled as membrane elements using the Finite Element Method.
Muscle material was modeled as a combination of fiber and an embedding matrix:
The fiber part was modelled according to Blemker et al.~\cite{Blemker2005a} using the following parameters: maximum isometric stress \mbox{$\sigma_{max}=\SI{6e5}{\newton\per\square\metre}$}, optimal fiber stretch \mbox{$\lambda_\text{ofl}=1.05$}, linear region starting stretch $\lambda^*$$=$$1.4$, and toe region dimensionless parameters $P_1$=$0.05$ and $P_2$=$6.6$.
The optimal fiber stretch ratio was set to 2 for the latissimus dorsi and teres major.\label{subsec:model}
A co-rotated Hooke’s law was used for the embedding matrix, with a Young's modulus of $E$=$\SI{1.5e3}{\newton\per\square\metre}$ and Poisson's ratio of $\nu$=$0.499$.
These membranes wrap the bones, relying on contact constraints. Contacts between muscles and bones use a penalty formulation with a stiffness of $10^5\,$\si{\newton\per\metre}.
Contacts between bones are enforced with hard constraints, while muscle-muscle contacts are omitted~\cite{Pean2020surface}.

The excitation of fibers in each muscle was quantified by a muscle activation parameter, normalized between 0 and 1.
In an inverse dynamics control scheme, the muscle activation parameters were determined by solving a local quadratic optimization problem at each simulation step~\cite{Stavness2012}.
The goal for the motion simulation was set to minimize the differences of positions and velocities between body markers from motion capture and corresponding simulation landmarks on the bone model.

\subsection{Subject-specific modelling}\label{sec:morphing}
In the current study, eating, washing, and combing motions from a publicly available shoulder movement dataset~\cite{Bolsterlee2014b} were utilized. 
This dataset contains motion capture of several upper limb landmarks: 4 
on the torso; 4 
on the scapula; 3 
on the clavicle; and 2 
on the humerus (with exact locations defined in~\cite{Wu2005}).
All motions for all 9 marker locations were transformed to a fixed thorax coordinate frame at all time instances by computing a rigid transformation of the thorax frame to its initial position at $t=0$.

The above landmarks were also annotated on our computational model.
Subject-specific tracking marker locations were transformed to our model coordinate frame and scale using the procedure detailed in appendix \ref{sec:apx}.
Accordingly, our model was transformed to fit the markers of Subject~4 from the dataset~\cite{Bolsterlee2014b}.
Thickness of muscle models was set to result in a volume value reported in the anthropomorphic measurements of the dataset. When such data was not available, the thickness was arbitrarily set to \SI{1}{\centi\metre}.
The mass of the forearm and hand was computed based on average anthropometric data~\cite{Leva1996} and represented as a point mass located in the middle between EM and EL.
Mass of the rest of the arm was accounted for as a point mass on the center of gravity of the humerus bone model.

\subsection{Simulation of Rotator Cuff Tears and Reverse Shoulder Arthroplasty}
\begin{figure}
	\def\svgwidth{\linewidth}
    \centering
    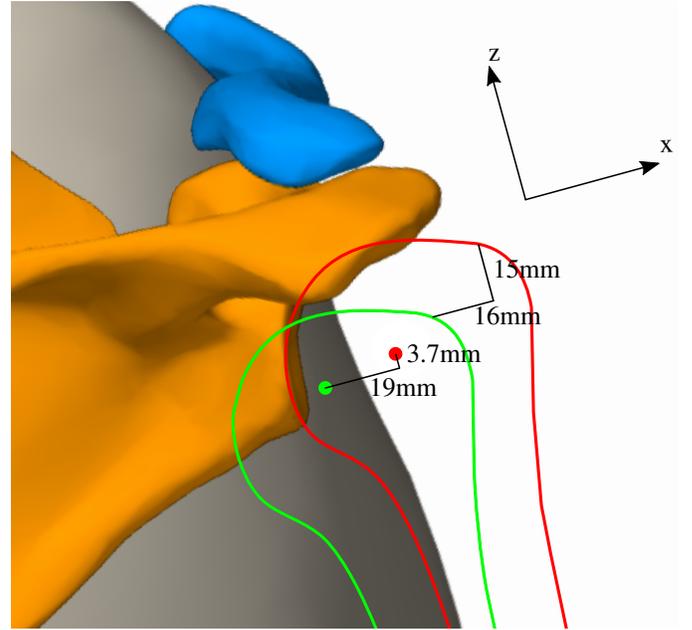%
    \caption{The effect of reverse prosthesis (green) modeled in our simulation by shifting the center of rotation (dots) of the glenohumeral joint~\cite{Boileau2005}, with red showing the intact case. Outlines of the humerus are shown as solid lines.}
    \label{fig:visualization}
\end{figure}

The intact situation with a well functioning rotator cuff (named ``intact") during anterior flexion was taken as the baseline.
Three levels of rotator cuff tears were considered: full tear of the supraspinatus (-ssp), combined full tear of the supraspinatus and subscapularis (-ssp-ssc), and combined full tear of the supraspinatus, infraspinatus, and teres minor (-ssp-isp-tmi), named superior, superior-anterior, and superior-posterior, respectively.
The tears were simulated by removing the respective muscles from the simulation, thus eliminating both their passive and active effects.
In this work, superior-anterior and superior-posterior tears are referred to as massive rotator cuff tears (MRCT).
The reverse implant was not modeled directly, but its effect on shoulder morphology and kinematics was incorporated by shifting the center of rotation of the glenohumeral joint.
Based on~\cite{Boileau2005} and as illustrated in \cref{fig:visualization}, we shifted the join center medially (towards the glenoid) by 19\,\si{\milli\meter} and inferiorly by 3.7\,\si{\milli\meter}, and translated the humerus medially (towards the glenoid) by 16\,\si{\milli\meter} and inferiorly by 15\,\si{\milli\meter}, which represents the average displacement of the humerus and of the center of rotation of the glenohumeral joint after the surgery.

\subsection{Model Validation}
In order to validate the model output, we compared peak JRF at 90\si{\degree} obtained from our intact model to in-vivo measurements obtained from instrumented prostheses, for the activities given in the work of Bergmann et al.~\cite{Bergmann2011}.
Herein we used the data for \emph{slow} forward flexion without 0.5kg dumbbell in the hand.
Furthermore, we compared the JRF throughout the entire anterior flexion motion to in-vivo measurements and to the Dutch Shoulder and Elbow Model (DSEM), a validated musculoskeletal model~\cite{Westerhoff2009,Nikooyan2011a}, for two subjects referred to as S1 and S2 in the work of Nikooyan et al.~\cite{Nikooyan2010}.

\subsection{Analysis}
The simulated glenohumeral JRF is decomposed into its compressive (JCF) and shear force components (JSF).
The JCF is defined as the component that is normal to the glenoid plane, labelled as the x-axis in \cref{fig:visualization}, while JSF lies in any direction within the glenoid plane.
Stability of the joint is warranted as long as the JRF vector points inside the glenoid. This is evaluated by ray-casting the JRF vector to a manual delineation of the glenoid outline extracted from the scapula mesh. If the JRF points outside of the glenoid boundaries, the joint is considered unstable.
Upward rotation of the scapula~\cite{Wu2005} was recorded throughout the movement, which is enabled by the free-hanging scapula and shoulder girdle muscles.
Range-of-motion was measured as the maximum reachable angle by the model during the simulation of the anterior flexion.

\section{Results}
\subsection{Validation}
\begin{figure}[t!]
	\def\svgwidth{\linewidth}
    \centering
    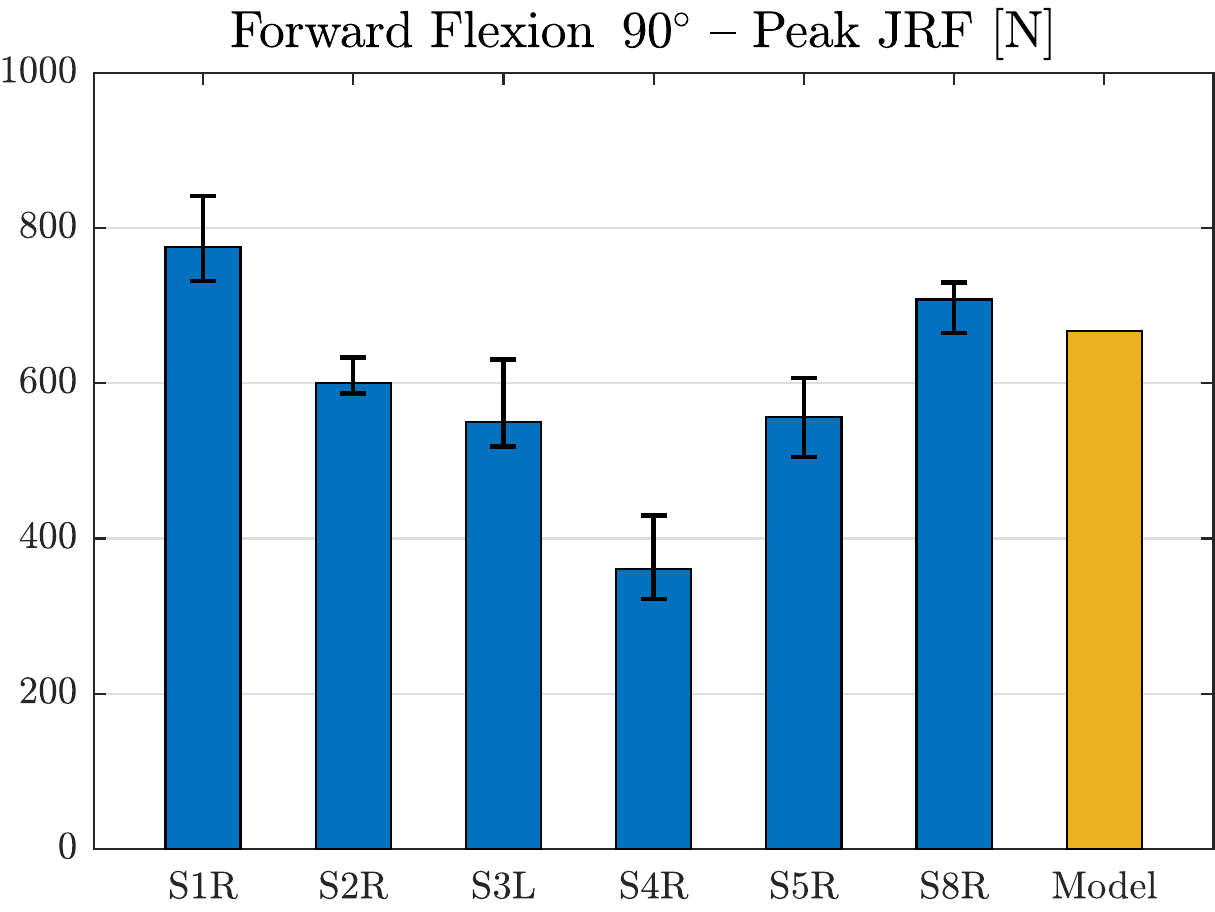%
    \caption{Comparison of peak JRF during anterior flexion up to 90\si{\degree} for slow forward flexion motion between the model and in-vivo measurements~\cite{Bergmann2011}.}
    \label{fig:jrf_orthoload}
    \bigskip
    \centering
    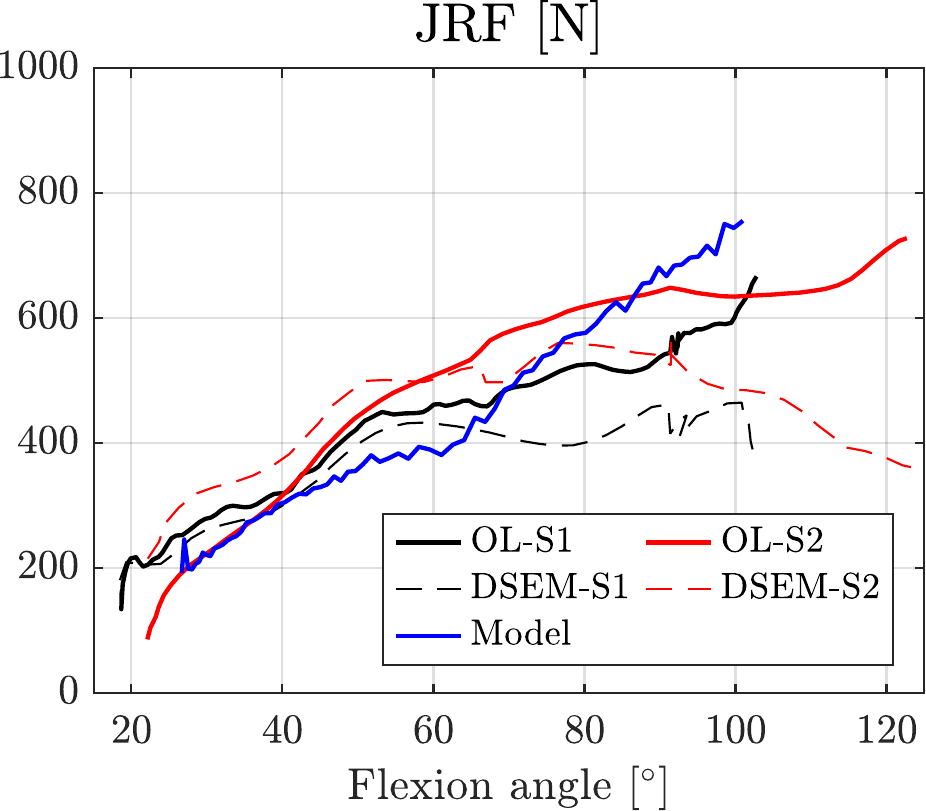%
    \caption{JRF during anterior flexion for our simulation, in-vivo measurements of subjects S1 and S2 from Orthoload~(OL)~\cite{Bergmann2008} and simulated subjects with DSEM~\cite{Nikooyan2010}.}
    \label{fig:jrf_nikooyan}
\end{figure}

The simulated peak JRF during anterior flexion up to 90\si{\degree} in comparison to in-vivo measurements is shown in \cref{fig:jrf_orthoload}.
Note that the subject numbering in \cref{fig:jrf_orthoload} is unrelated to those from the motion capture dataset that we used to drive our model in \cref{sec:morphing}.
The predicted peak JRF lies within the range covered by in-vivo measurements.
The JRF during anterior flexion shows a similar trend as OL-S1 up to 60\si{\degree}, however past this point, the two models clearly behave differently (\cref{fig:jrf_nikooyan}).
While the JRF tends to decrease for the DSEM model after 60\si{\degree}, our simulation presents a similar trend to in-vivo measurements with the JRF continuing to increase at flexion angles above 90\si{\degree}.

\subsection{Range-of-motion}
Range-of-motion improved with a reverse implant for cases involving a RCT, see \cref{tab:rom}.
The improvement was particularly noticeable for the superior-posterior MRCT (-ssp-isp-tmi), which increased from 51\si{\degree} to 98\si{\degree}.

\begin{table}
\centering
\caption{Range-of-motion during anterior flexion with (+rev) and without reverse prosthesis.}
\label{tab:rom}
\begin{tabular}{ccccc}
\toprule
{[\si{\degree}]} & intact & -ssp & -ssp-ssc & -ssp-isp-tmi\\
\midrule
                    & 106  & 84   & 90       & 51\\
+rev                & 107  & 96   & 101      & 98\\
\bottomrule
\end{tabular}
\end{table}

\subsection{Flexion with tears of the Rotator Cuff}
\begin{figure*}
	\def\svgwidth{\textwidth}
    \centering
    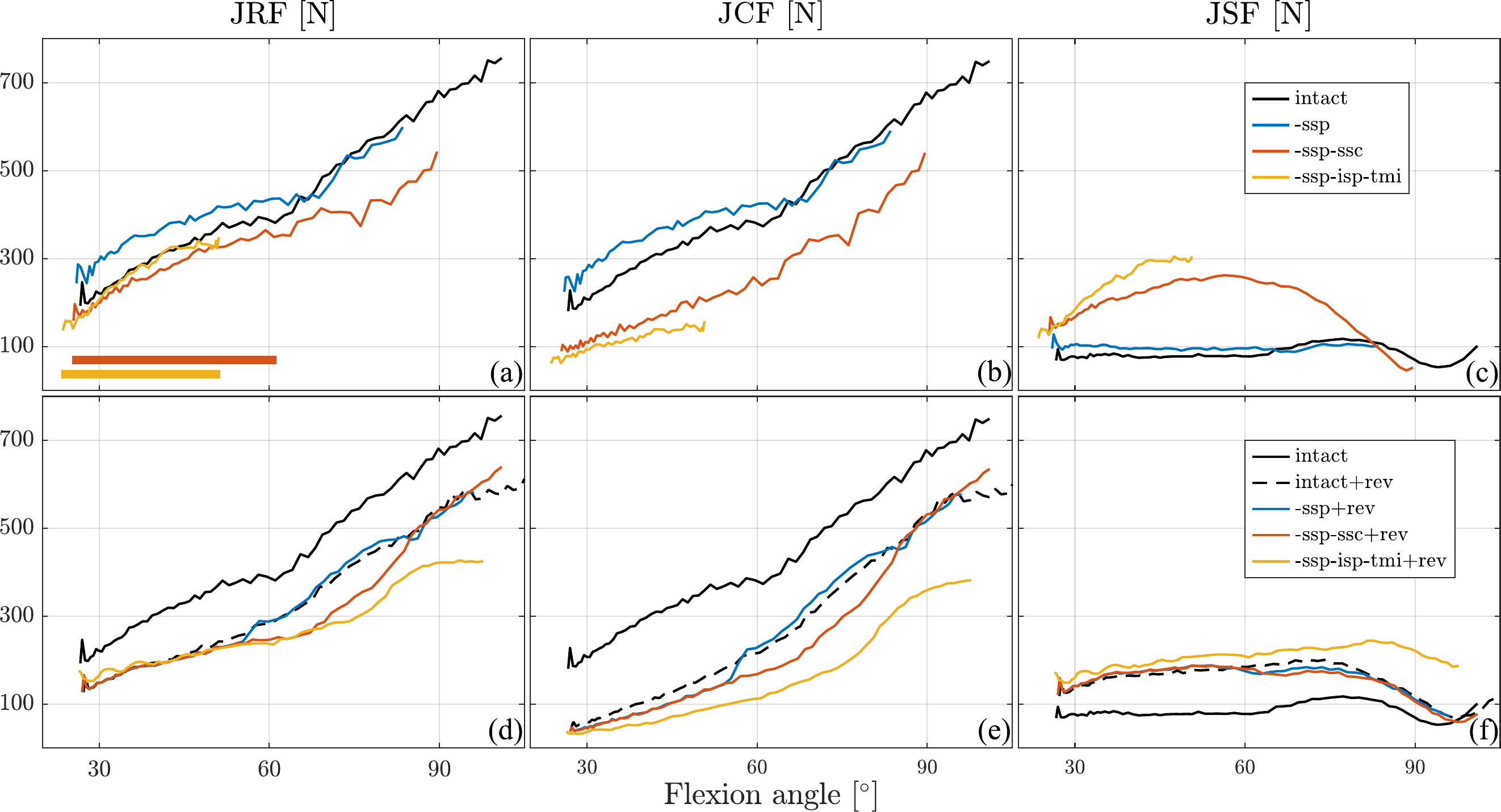%
    \caption{Force magnitude in the glenohumeral joint during anterior flexion~(a,d), detailed into its compressive ~(b,e) and shear~(c,f) components, for different rotator cuff tear conditions ~(a,b,c) and with reverse implant~(d,e,f). The range when JRF points outside the glenoid is shown as a horizontal bar in the JRF graph with the  color corresponding to muscle condition in~(a).}
    \label{fig:flexion_force}
\end{figure*}

With a tear of the supraspinatus(-ssp), the JRF stayed within the glenoid during the full flexion motion.
Compared to the intact joint, an average JRF increase of 17\%  was observed between 25 to 65 degrees of flexion, with a peak reaching up to 34\% (\cref{fig:flexion_force}a).
This increase occurs both for the shear and compressive components, increasing by 23 and 17\% respectively (\cref{fig:flexion_force}b and \cref{fig:flexion_force}c).

\begin{figure}[t]
	\def\svgwidth{\linewidth}
    \centering
    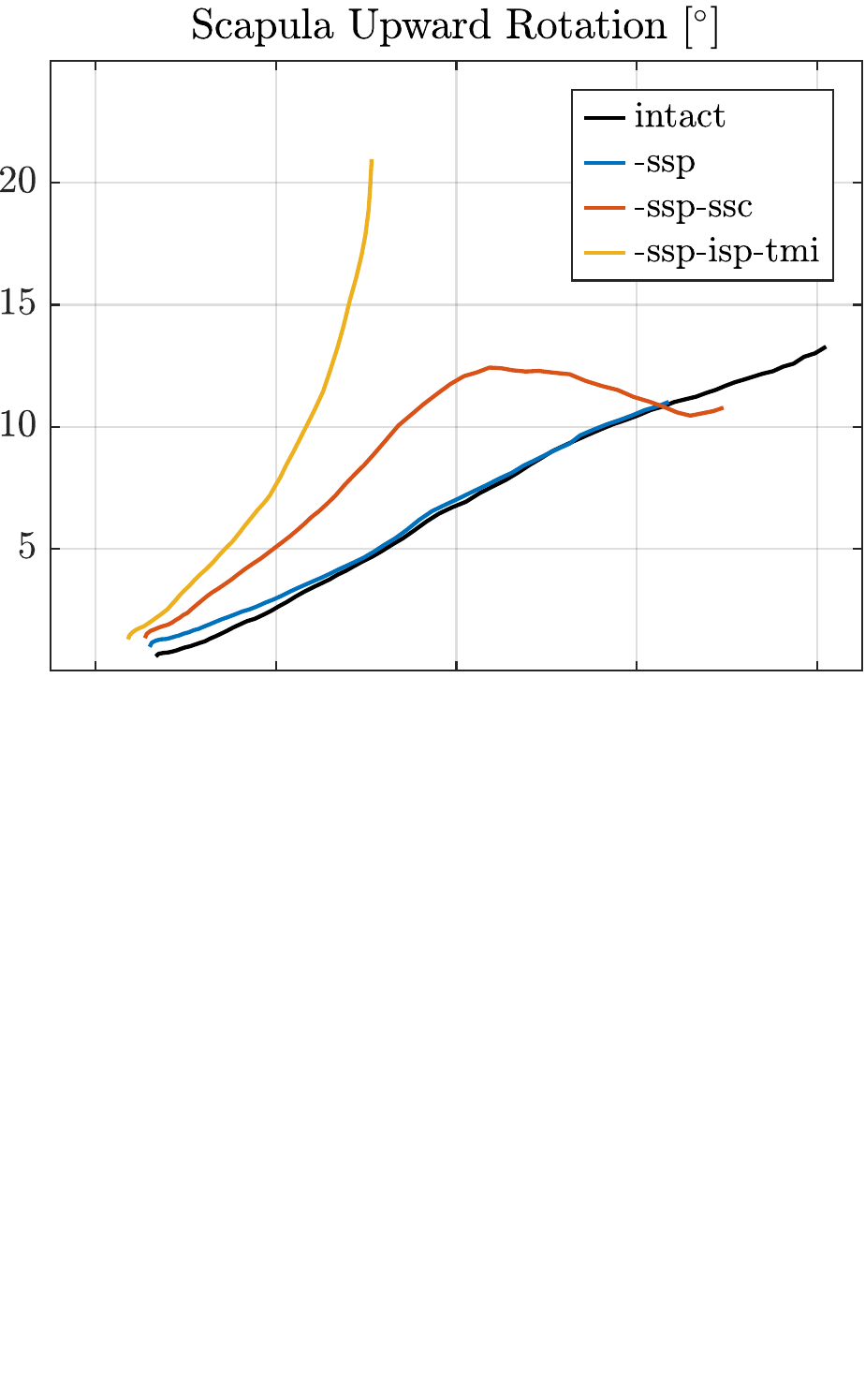    \caption{Upward rotation of the scapula for the different rotator cuff tears~(a) and reverse arthroplasty~(b) conditions. The intact anatomy baseline is shown as a solid black line.}
    \label{fig:flexion_scapula_rot}
\end{figure}

With MRCT~(-ssp-ssc and -ssp-isp-tmi), we observe an important reduction of JCF and a significant increase of JSF compared to the intact joint (\cref{fig:flexion_force}b and \cref{fig:flexion_force}c).
The JRF is directed outside the glenoid up to 66\si{\degree} for superior-anterior tears~(-ssp-ssc) and throughout the whole motion for superior-posterior tears~(-ssp-isp-tmi)(\cref{fig:flexion_force}a). 
In practice, this may lead to joint instability and dislocation in the absence of additional structures such as ligaments or capsules.
With MRCT, when the JRF points outside the glenoid, i.e.\ below $61\si{\degree}$ (\cref{fig:flexion_force}a), a compensation by the scapula is observed, leading to 16\si{\degree} higher upward rotation than in the intact case (\cref{fig:flexion_scapula_rot}a). 
Once stability is restored, beyond $61\si{\degree}$ for superior-anterior tears~(-ssp-ssc), rotation of the scapula returns toward~(intact) baseline values.

\subsection{Flexion with RSA}
The JCF for an intact RC (intact+rev) was higher than that of RSA with superior-anterior tears (-ssp-ssc+rev) for up to 85\si{\degree} of anterior flexion, and over the entire motion for RSA with superior-posterior tears (-ssp-isp-tmi+rev) (\cref{fig:flexion_force}e).
Over the given ranges, JCF for the intact RC was on average 26\% higher than that for the superior-anterior tear with a reverse prosthesis~(-ssp-ssc+rev), and 76\% higher than that for the superior-posterior tear with a reverse prosthesis~(-ssp-isp-tmi+rev).
The JSF of intact was similar to the superior-anterior tear with reverse prosthesis case (-ssp-ssc+rev) (\cref{fig:flexion_force}f).

The JSF of the superior tear with reverse(-ssp+rev) is similar to that of the superior-anterior tear condition with reverse (-ssp-ssc+rev). However, JCF is higher for superior-anterior tears with reverse implant(-ssp-ssc+rev) in a range of 55 to 85\si{\degree} of anterior flexion, and over the entire motion for superior-posterior tears(-ssp-isp-tmi+rev), see \cref{fig:flexion_force}e.
Over the given ranges, JCF averaged 31\% higher than for the superior-anterior tear with a reverse prosthesis(-ssp-ssc+rev), and 65\% higher than that for the superior-posterior tear with a reverse prosthesis(-ssp-isp-tmi+rev).

The JRF with a reverse prosthesis is generally lower than in the intact case, with a drop of 32\% for the superior-anterior tear~(-ssp-ssc+rev) and 37\% for the superior-posterior MCRT~(-ssp-isp-tmi+rev) (\cref{tab:jrf}).
Upward rotation of the scapula was higher with RSA than that for the intact case, up to 80\si{\degree} improvement in anterior elevation, and throughout the whole motion for the superior-posterior tear condition (-ssp-isp-tmi+rev) (\cref{fig:flexion_scapula_rot}).

\begin{table}
\centering
\caption{Mean difference in JRF with respect to intact case during anterior flexion with (+rev) and without prosthesis.}
\label{tab:jrf}
\begin{tabular}{ccccc}
\toprule
{[\%]} & intact & -ssp & -ssp-ssc & -ssp-isp-tmi\\
\midrule
                    & 0  & 11 & -15 & -5       \\
+rev                &-28  & -28 & -32 & -37      \\
\bottomrule
\end{tabular}
\end{table}

\section{Discussion}
A major strength of this study is that our results favorably corroborate JRF measured in-vivo as well as clinical observations with a reverse prosthesis (restoration of function and scapular compensation, with and without a reverse implant), as described below.
JRF generated by our model match well with in-vivo measurements~\cite{Nikooyan2010,Bergmann2011}.
In contrast to the DSEM, JRF in our model do not decrease at high flexion angles,
allowing for the analysis of a larger range of motion and potentially indicating the validity of our model being comparatively higher.
One model in the literature~\cite{Favre2009} reported similar JRF increase, but this result was later shown to be dependent on the chosen anatomical dataset~\cite{Favre2012}, highlighting the influence of  muscle arrangements.
In the current model, this behavior may also be attributed to the specificity of the chosen anatomy, or to the modelling of muscles as surfaces without via-points, which improves wrapping around bones and creates physiologically more realistic models~\cite{Webb2014,Pean2020surface,Aurbach2020}.
While our simulated JRF presents a steeper slope than in-vivo recordings at flexion angles over 65\si{\degree}, it overall remains close to the range covered by the in-vivo data while demonstrating a reasonable trend.

The simulated reduction of JRF after a reverse prosthesis for MRCT (\cref{tab:jrf}) corroborated previous studies~\cite{Terrier2008,Henninger2012}, where a 30\% JRF reduction was predicted during abduction with a reverse implant compared to the intact anatomy.
Our model shows a range-of-motion improvement for all rotator cuff tear conditions with a reverse implant, a finding in agreement with the clinical literature~\cite{Favard2011,Nolan2011,Petrillo2017,Lindbloom2019}.
Furthermore, the predicted improvement from 51 to 98\si{\degree} for a superior-posterior MRCT  (\cref{tab:rom}) is in line with reported clinical values of recovery from an average of 51\si{\degree} preoperative to 124\si{\degree} post-operative~\cite{Petrillo2017}.

In clinical practice, compensation by the scapula is often observed in MRCT cases~\cite{Familiari2018}.
Our model was also able to replicate this clinical observation (\cref{fig:flexion_scapula_rot}).
Additionally, increased  scapular upward rotation with a reverse implant reported in~\cite{Kim2012how,Walker2015,Lee2016three} was also reproduced by our model, particularly for the superior-posterior tear.
Scapular kinematics was not imposed in our model, as commonly done in shoulder modeling~\cite{Holzbaur2005,Dickerson2007,Nikooyan2011a}: Instead, the scapula was actively positioned by the muscles of the shoulder girdle. In the presence of a MRCT, it was more effective for the controller of our model to actively rotate the scapula, rather than imposing high forces from ineffective muscles to achieve humeral elevation.

The supraspinatus is a key muscle in early abduction and flexion~\cite{Kincaid2006,Reed2013,DeWitte2014} and its tear can be compensated by the upper portions of the subscapularis and infraspinatus~\cite{Jost2003}. 
It leads to a stronger co-contraction of these two muscles compared to the intact case (activation is superior by 7.0 and 9.6\% in average for the subscapularis and infraspinatus, respectively).
The lower efficiency of these muscle segments compared to the supraspinatus to perform this function can explain the visibly higher JCF (\cref{fig:flexion_force}b).
The tear of an additional rotator cuff muscle likely disturbs the anterior-posterior balance around the humeral head, and function is lost due to the impossibility of co-contraction.
As the rotator cuff tear progresses, the glenohumeral joint loses its main stabilizers, which provides the main compressive action on the joint~\cite{Parsons2002}, resulting in severe reduction of JCF and leading to joint instability (\cref{fig:flexion_force}b-c).
Implanting a reverse prosthesis enables to convert the shear action of the deltoid on the joint into usable torque for the movement, which is one of the main design motivations of the reverse prosthesis~\cite{Boileau2005}.

Implanting a reverse shoulder prosthesis with a functional rotator cuff resulted in a JCF increase of up to $1.4$ and $2.1$ times, when compared, respectively, to a superior-anterior tear with prosthesis and a superior-posterior tear with prosthesis (\cref{fig:flexion_force}e). 
While this may have a positive impact on joint stability, higher compression forces might increase the risk of wear and implant fracture.

Limitations of the model include an overly constrained glenohumeral joint, as the kinematics of a ball-socket joint does not allow translation of the humerus within the glenoid cavity~\cite{Favre2012,Quental2016a}.
Our inverse dynamics control algorithm does not include active stabilisation for constraining JRF within the glenoid cavity, leading potentially to an under-estimation of JRF~\cite{Favre2005} in the MRCT cases (-ssp-ssc and -ssp-isp-tmi), whereby the joint might be unstable.
The simulations were performed for a single anatomy, although subject variability is assumed to play a large role. Further studies are required to confirm our finding for a larger population. Especially, the muscle rest length is crucial for determining passive forces by the muscles and inter-subject variations may have a large effect on the range-of-motion.

\section{Conclusions}
Observations from our model compare well with in-vivo measurements, while also corroborating several previously published clinical and biomechanical observations.
We found that with a simulated reverse implant, a higher glenohumeral compressive joint force is applied with a mostly-functional rotator cuff compared to a MRCT, while the shear force component is reduced or kept similar. While this phenomenon may improve joint stability, it may increase the risk for other failure modes such as fracture or polyethylene wear of the reverse prosthesis.
This may be of importance when considering RSA for relatively younger patients.

\bibliographystyle{abbrvnat}
\bibliography{arxiv}

\appendices
\section{Subject-specific transformation of tracked motion to our model coordinates}
\label{sec:apx}
Our goal is to be able to utilize motion tracking data from an arbitrary subject (in this paper the dataset of Bolsterlee) as functional simulation target in our specific shoulder model.
Since the dimensions of our model and the tracked subject may not match, this is not a trivial problem.
We propose herein to ``morph'' the tracked motion (i.e.\ a set of anatomical landmark locations in time) to our model, in the following procedural way to preserve most physical and anatomical conditions.
Below we use the tracking landmark abbreviations following the ISB recommendations~\cite{Wu2005}.

The scapula is a major component of shoulder motion, and hence both our model and the motion capture.
Accordingly, we use a scapula-centered transformation approach.
To align the model scales, we first register our entire rest-state model to the subject-specific scapular landmarks in the initial tracking instance with a Procrustes transformation. 
Then, to align the thorax, we rotate the tracked thorax landmarks around the center point defined by TS, AI, and GL (center of a sphere fitted to the glenoid in our model), with a rotation to vertically align C7 and T8.
The thorax is then rotated around the vertical axis to anteriorly align C7 and IJ.
The clavicle is then registered using the SC, ACd, ACv landmarks with a Procrustes transform.
The humerus is rigidly rotated around GH (center of a sphere fitted on the humeral head in our model) to align EM and EL, for a least-square fit between the humerus landmarks in the subject-specific tracked motion and on our model.

The static registration above ensures a well-posed fitting of our model to the dataset, while preserving the shapes of anatomical structures.
However, any jitter or errors in the tracking or anatomical differences of the subject may still cause the transformed points to reach locations infeasible by our model.
To prevent this, in a pre-processing step we smooth the tracked positions over time using landmark positions reachable by an inverse simulation of our model.
The kinematic chain of our skeletal model can be seen in~\cite{Pean2020surface}-Figure\,3.
In our simulations, to provide the tracked scapula as a soft constraint, the scapular landmarks on our model are duplicated, with one set attached on our bone model and the other set registered to the tracking according to the static procedure above for all time steps.
During simulation, springs between these two landmark sets allow for our scapula model to follow the transformed version of the tracked scapular landmarks.
The rotation of the humerus is resolved at each time step with the static procedure above.
The positions of the landmarks attached onto the bones are recorded and used as desired target input to the inverse dynamic musculoskeletal simulation.
\end{document}

%% file: fig/translation_clipped_hor_2_svg-tex.pdf_tex
\begingroup%
  \makeatletter%
  \providecommand\color[2][]{%
    \errmessage{(Inkscape) Color is used for the text in Inkscape, but the package 'color.sty' is not loaded}%
    \renewcommand\color[2][]{}%
  }%
  \providecommand\transparent[1]{%
    \errmessage{(Inkscape) Transparency is used (non-zero) for the text in Inkscape, but the package 'transparent.sty' is not loaded}%
    \renewcommand\transparent[1]{}%
  }%
  \providecommand\rotatebox[2]{#2}%
  \newcommand*\fsize{\dimexpr\f@size pt\relax}%
  \newcommand*\lineheight[1]{\fontsize{\fsize}{#1\fsize}\selectfont}%
  \ifx\svgwidth\undefined%
    \setlength{\unitlength}{327.94606409bp}%
    \ifx\svgscale\undefined%
      \relax%
    \else%
      \setlength{\unitlength}{\unitlength * \real{\svgscale}}%
    \fi%
  \else%
    \setlength{\unitlength}{\svgwidth}%
  \fi%
  \global\let\svgwidth\undefined%
  \global\let\svgscale\undefined%
  \makeatother%
  \begin{picture}(1,0.94117653)%
    \lineheight{1}%
    \setlength\tabcolsep{0pt}%
    \put(0,0){\includegraphics[width=\unitlength,page=1]{translation_clipped_hor_2_svg-tex.pdf}}%
    \put(0.97300358,0.71504636){\color[rgb]{0,0,0}\rotatebox{2.267296}{\makebox(0,0)[lt]{\lineheight{1.25}\smash{\begin{tabular}[t]{l}x\end{tabular}}}}}%
    \put(0.715677,0.85304065){\color[rgb]{0,0,0}\rotatebox{2.267296}{\makebox(0,0)[lt]{\lineheight{1.25}\smash{\begin{tabular}[t]{l}z\end{tabular}}}}}%
    \put(0,0){\includegraphics[width=\unitlength,page=2]{translation_clipped_hor_2_svg-tex.pdf}}%
    \put(0.69312758,0.45680999){\color[rgb]{0,0,0}\makebox(0,0)[lt]{\lineheight{1.25}\smash{\begin{tabular}[t]{l}16mm\end{tabular}}}}%
    \put(0,0){\includegraphics[width=\unitlength,page=3]{translation_clipped_hor_2_svg-tex.pdf}}%
    \put(0.72211079,0.52604099){\color[rgb]{0,0,0}\makebox(0,0)[lt]{\lineheight{1.25}\smash{\begin{tabular}[t]{l}15mm\end{tabular}}}}%
    \put(0.59306544,0.39988199){\color[rgb]{0,0,0}\makebox(0,0)[lt]{\lineheight{1.25}\smash{\begin{tabular}[t]{l}3.7mm\end{tabular}}}}%
    \put(0.53651637,0.34849225){\color[rgb]{0,0,0}\makebox(0,0)[lt]{\lineheight{1.25}\smash{\begin{tabular}[t]{l}19mm\end{tabular}}}}%
  \end{picture}%
\endgroup%

%% file: fig/flexion_bars_slow0kg_svg-tex.pdf_tex
\begingroup%
  \makeatletter%
  \providecommand\color[2][]{%
    \errmessage{(Inkscape) Color is used for the text in Inkscape, but the package 'color.sty' is not loaded}%
    \renewcommand\color[2][]{}%
  }%
  \providecommand\transparent[1]{%
    \errmessage{(Inkscape) Transparency is used (non-zero) for the text in Inkscape, but the package 'transparent.sty' is not loaded}%
    \renewcommand\transparent[1]{}%
  }%
  \providecommand\rotatebox[2]{#2}%
  \newcommand*\fsize{\dimexpr\f@size pt\relax}%
  \newcommand*\lineheight[1]{\fontsize{\fsize}{#1\fsize}\selectfont}%
  \ifx\svgwidth\undefined%
    \setlength{\unitlength}{349.8124875bp}%
    \ifx\svgscale\undefined%
      \relax%
    \else%
      \setlength{\unitlength}{\unitlength * \real{\svgscale}}%
    \fi%
  \else%
    \setlength{\unitlength}{\svgwidth}%
  \fi%
  \global\let\svgwidth\undefined%
  \global\let\svgscale\undefined%
  \makeatother%
  \begin{picture}(1,0.75040203)%
    \lineheight{1}%
    \setlength\tabcolsep{0pt}%
    \put(0,0){\includegraphics[width=\unitlength,page=1]{flexion_bars_slow0kg_svg-tex.pdf}}%
  \end{picture}%
\endgroup%

%% file: fig/flexion_orthoload_svg-tex.pdf_tex
\begingroup%
  \makeatletter%
  \providecommand\color[2][]{%
    \errmessage{(Inkscape) Color is used for the text in Inkscape, but the package 'color.sty' is not loaded}%
    \renewcommand\color[2][]{}%
  }%
  \providecommand\transparent[1]{%
    \errmessage{(Inkscape) Transparency is used (non-zero) for the text in Inkscape, but the package 'transparent.sty' is not loaded}%
    \renewcommand\transparent[1]{}%
  }%
  \providecommand\rotatebox[2]{#2}%
  \newcommand*\fsize{\dimexpr\f@size pt\relax}%
  \newcommand*\lineheight[1]{\fontsize{\fsize}{#1\fsize}\selectfont}%
  \ifx\svgwidth\undefined%
    \setlength{\unitlength}{266.5195875bp}%
    \ifx\svgscale\undefined%
      \relax%
    \else%
      \setlength{\unitlength}{\unitlength * \real{\svgscale}}%
    \fi%
  \else%
    \setlength{\unitlength}{\svgwidth}%
  \fi%
  \global\let\svgwidth\undefined%
  \global\let\svgscale\undefined%
  \makeatother%
  \begin{picture}(1,0.87529458)%
    \lineheight{1}%
    \setlength\tabcolsep{0pt}%
    \put(0,0){\includegraphics[width=\unitlength,page=1]{flexion_orthoload_svg-tex.pdf}}%
  \end{picture}%
\endgroup%

%% file: fig/flexion_hor_svg-tex.pdf_tex
\begingroup%
  \makeatletter%
  \providecommand\color[2][]{%
    \errmessage{(Inkscape) Color is used for the text in Inkscape, but the package 'color.sty' is not loaded}%
    \renewcommand\color[2][]{}%
  }%
  \providecommand\transparent[1]{%
    \errmessage{(Inkscape) Transparency is used (non-zero) for the text in Inkscape, but the package 'transparent.sty' is not loaded}%
    \renewcommand\transparent[1]{}%
  }%
  \providecommand\rotatebox[2]{#2}%
  \newcommand*\fsize{\dimexpr\f@size pt\relax}%
  \newcommand*\lineheight[1]{\fontsize{\fsize}{#1\fsize}\selectfont}%
  \ifx\svgwidth\undefined%
    \setlength{\unitlength}{761.44309395bp}%
    \ifx\svgscale\undefined%
      \relax%
    \else%
      \setlength{\unitlength}{\unitlength * \real{\svgscale}}%
    \fi%
  \else%
    \setlength{\unitlength}{\svgwidth}%
  \fi%
  \global\let\svgwidth\undefined%
  \global\let\svgscale\undefined%
  \makeatother%
  \begin{picture}(1,0.54103427)%
    \lineheight{1}%
    \setlength\tabcolsep{0pt}%
    \put(0,0){\includegraphics[width=\unitlength,page=1]{flexion_hor_svg-tex.pdf}}%
  \end{picture}%
\endgroup%

%% file: fig/flexion_scapula_rot_combined_svg-tex.pdf_tex
\begingroup%
  \makeatletter%
  \providecommand\color[2][]{%
    \errmessage{(Inkscape) Color is used for the text in Inkscape, but the package 'color.sty' is not loaded}%
    \renewcommand\color[2][]{}%
  }%
  \providecommand\transparent[1]{%
    \errmessage{(Inkscape) Transparency is used (non-zero) for the text in Inkscape, but the package 'transparent.sty' is not loaded}%
    \renewcommand\transparent[1]{}%
  }%
  \providecommand\rotatebox[2]{#2}%
  \newcommand*\fsize{\dimexpr\f@size pt\relax}%
  \newcommand*\lineheight[1]{\fontsize{\fsize}{#1\fsize}\selectfont}%
  \ifx\svgwidth\undefined%
    \setlength{\unitlength}{266.0274bp}%
    \ifx\svgscale\undefined%
      \relax%
    \else%
      \setlength{\unitlength}{\unitlength * \real{\svgscale}}%
    \fi%
  \else%
    \setlength{\unitlength}{\svgwidth}%
  \fi%
  \global\let\svgwidth\undefined%
  \global\let\svgscale\undefined%
  \makeatother%
  \begin{picture}(1,1.61768092)%
    \lineheight{1}%
    \setlength\tabcolsep{0pt}%
    \put(0,0){\includegraphics[width=\unitlength,page=1]{flexion_scapula_rot_combined_svg-tex.pdf}}%
    \put(0.86283165,0.8754571){\color[rgb]{0,0,0}\makebox(0,0)[lt]{\lineheight{1.25}\smash{\begin{tabular}[t]{l}(a)\end{tabular}}}}%
    \put(0.85973484,0.15936545){\color[rgb]{0,0,0}\makebox(0,0)[lt]{\lineheight{1.25}\smash{\begin{tabular}[t]{l}(b)\end{tabular}}}}%
    \put(0,0){\includegraphics[width=\unitlength,page=2]{flexion_scapula_rot_combined_svg-tex.pdf}}%
  \end{picture}%
\endgroup%